\documentclass[11pt,twoside]{article}
\usepackage{asp2004}
\usepackage{psfig}
\usepackage{epsf}
\usepackage{graphics}
\usepackage{lscape}
\def\edcomment#1{\iffalse\marginpar{\raggedright\sl#1\/}\else\relax\fi}
\reversemarginpar

\begin{document}
\title{Magnetorotational supernova simulations}
\author{S.G.Moiseenko$^1$, G.S.Bisnovatyi-Kogan$^1$, N.V.Ardeljan$^2$}
\affil{$^1 $Space Research Institute, Profsoyuznaya str. 84/32
Moscow 117997, Russia, moiseenko@iki.rssi.ru, gkogan@iki.rssi.ru
\\
$^2$Department of Computational Mathematics and Cybernetics,
Moscow State University, Vorobjevy Gory, Moscow B-234 Russia,
ardel@cs.msu.su}

\begin{abstract}
We present 2D results of simulations of the magnetorotational core
collapsed supernova. For the first time we obtain strong explosion
for the core collapsed supernova. In 2D approximation we show that
amplification of the toroidal magnetic field due to the
differential rotation leads to the formation of MHD shockwave,
which produces supernova explosion. The amounts of the ejected
mass $0.1M_\odot$ and energy  $\sim 0.5\div0.6 \cdot 10^{51}$ergs
can explain the energy output for supernova type II or type Ib/c
explosions. The shape of the explosion is qualitatively depends on
the initial configuration of the magnetic field, and may form
strong ejection neat the equatorial plane, or produce mildly
collimated jets. Our simulation show that during the evolution of
the magnetic field the magnetorotational instability appears and
leads to exponential growth of the magnetic field strength.
\end{abstract}
\thispagestyle{plain}

\section{Introduction}
The reliable explanation of the mechanism of core collapsed
supernova is still open question for the modern astrophysics. We
describe the results of 2D numerical simulation of
magnetorotational mechanism for core collapsed supernova. The main
idea of the magnetorotational mechanism for the core collapsed
supernova was suggested by \citet{bk1970}. After the core collapse
central parts of the star rotate differentially. In the presence
of initial poloidal magnetic field toroidal component of the
magnetic field appears and amplifies with time. When the pressure
of the magnetic field becomes comparable with the gas pressure the
compression wave arise and moves along steeply decreasing density
profile. In a short time this compression wave transforms to the
MHD fast shock wave and produces supernova explosion. First 2D
simulations of the magnetorotational mechanism were made by
\citet{lw1970} with unrealistically large initial magnetic field.
2D simulations of the similar problem were given in the papers of
\citet{ohnishi} and \citet{symbalisty}. The magnetorotational
processes with very large initial magnetic fields (typical for
magnetars) were recently simulated numerically in the papers of
\citet{kotake} and \citet{yamada}. The 1D simulations of the
magnetorotational supernova were presented in the papers by
\citet{bk76}, \citet{abkp}. In the last paper the simulations were
performed for a wide range of the initial values of magnetic
field. 2D simulations of the collapse of the magnetized rotating
protostellar cloud are presented in the paper by \citet{abkm}.
\begin{figure}
\centerline{\includegraphics{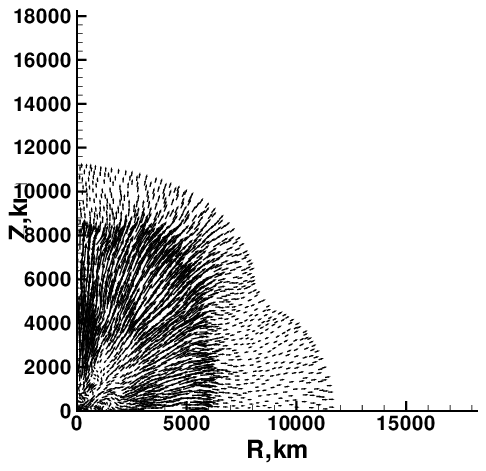},\includegraphics{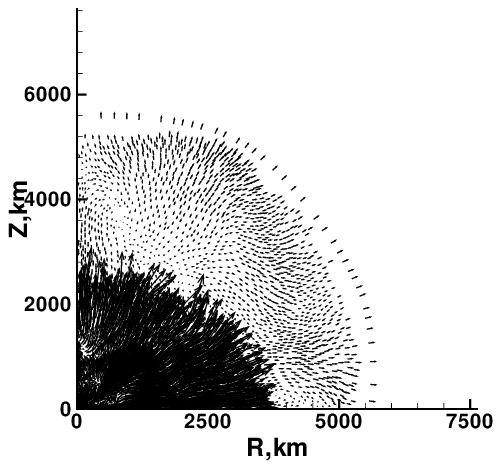}}
 \caption{Velocity field for $t=1.06$s after beginning the evolution of the toroidal magnetic field component for the
 {\it dipole}-like initial magnetic field (left plot), velocity field for $t=0.2$s after beginning the evolution of the toroidal magnetic field component for the
 {\it quadrupole}-like initial magnetic field (right plot).}
  \label{veldipquad}
\end{figure}
\begin{figure}
\centerline{\includegraphics{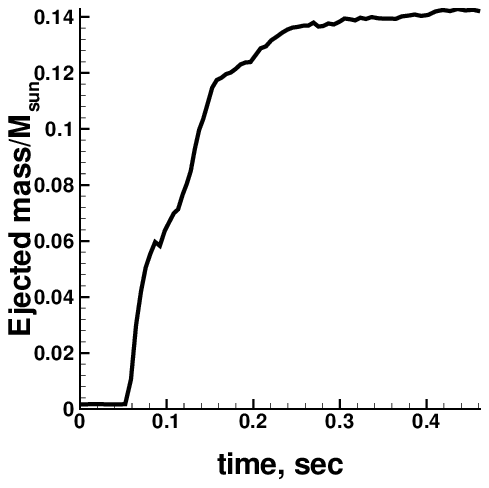}\includegraphics{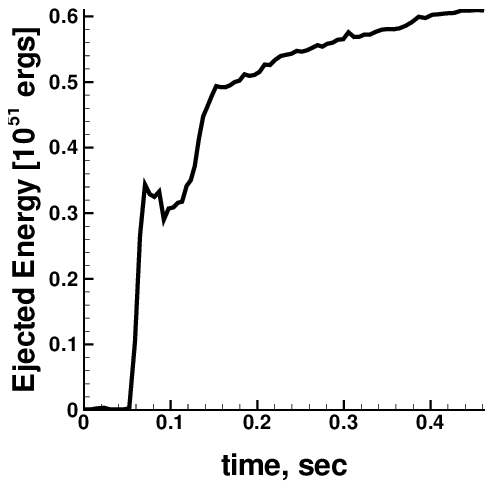}}
 \caption{Time evolution for of the ejected mass (in relation to $M_\odot$) and energy
for the {\it quadrupole}-like initial magnetic field,
$\alpha=10^{-6}$.}
  \label{masenq}
\end{figure}
\begin{figure}
\centerline{\includegraphics{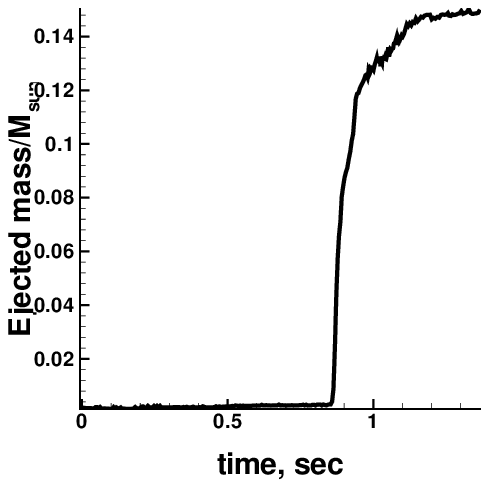}\includegraphics{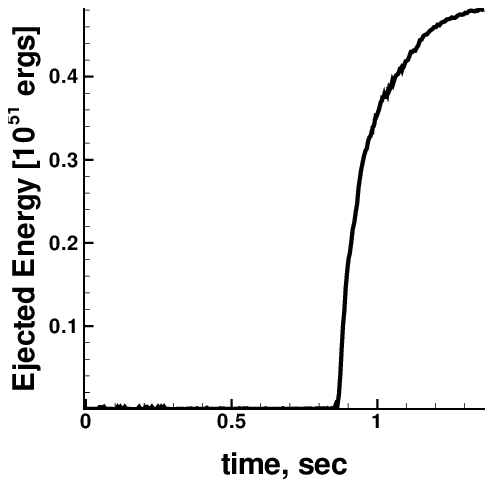}}
 \caption{Time evolution of the ejected mass (in relation to  $M_\odot$) and energy
for the {\it dipole}-like initial magnetic field,
$\alpha=10^{-6}$.}
  \label{masend}
\end{figure}
\section{Formulation of the problem}
For the numerical simulation of the magnetorotational supernova we
use set of MHD equations with selfgravitation and with infinite
conductivity in cylindrical coordinates. The equations of state
and neutrino losses are described in details in the paper of
\citet{abkkm}. Axial symmetry ($\frac \partial {\partial
\varphi}=0$, $r\geq 0$) and symmetry to the equatorial plane
($z\geq0$) are assumed. For the simulations we used specially
developed implicit numerical method based on the Lagrangian
operator-difference scheme on triangular grid of variable
structure, see \citet{arkos}.

\section{Dipole-like and quadrupole-like initial magnetic fields}
For the simulations we used initial dipole-like and quadrupole-like initial magnetic fields.
Magnetorotational explosion with the initial dipole-like magnetic field has the shape of mildly
collimated jet (Figure \ref{veldipquad}, left plot).  This jet can be collimated later when the shock wave
 pass through the envelope of the star. Initial quad\-ru\-po\-le-like poloidal
magnetic field leads to the explosion which develops preferably near equatorial plane
(Figure \ref{veldipquad}, right plot).

In the both cases (dipole and quadrupole initial poloidal magnetic
field) the amplification of the toroidal magnetic field leads to
the formation of the compression wave, which moves along steeply
decreasing density profile, and soon it transforms to the fast MHD
shock wave.

For the quadrupole initial magnetic field the evolution of the magnetorotational explosion takes place
much faster then for the dipole field (Figure\ref{masenq}, \ref{masend}).

The simulations of the magnetorotational explosion for the initial
quad\-ru\-po\-le and dipole magnetic field lead to the ejection of
about $0.6\cdot 10^{51}$ergs of energy and about $0.14M_\odot$ of
mass. The amount of the ejected mass will increase with the
evolution of the MHD shock.

\begin{figure}
\centerline{\includegraphics{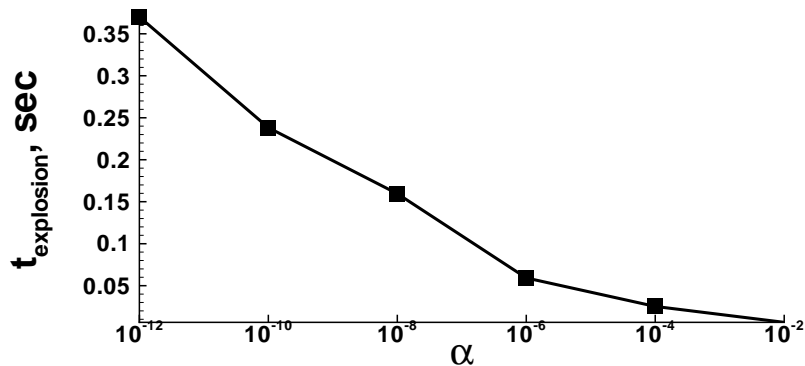}}
 \caption{Dependence of the explosion time from  $\alpha=\frac{E_{mag0}}{E_{grav}}$.}
 \label{texpl}
\end{figure}
\section{Magnetorotational instability in the magnetorotational mechanism}
Let $\alpha$ be the ratio of the initial poloidal magnetic energy
of the star and its gravitational energy at the moment of the
"turning on" magnetic field. $\alpha=\frac{E_mag0}{Egrav}$. We
have made simulations for the initial quadrupole-like magnetic
field for the wide range of the initial values of the magnetic
energy ($\alpha=10^{-2}\div 10^{-12}$). The values of the
explosion energy and ejected mass are approximately the same for
all calculated variants. In 1D simulations of the
magnetorotational mechanism by \citet{abkp} it was shown that
evolution time of the magnetorotational explosion depends on
$\alpha$ in the following way $t_{explosion} \sim
\frac{1}{\sqrt{\alpha}}$. In 2D simulations the situation as
qualitatively different. Due to the appearing and evolution of the
magnetorotational instability, see \citet{dungey}, \citet{tayler},
\citet{spruit}, \citet{akiyama}, the time from the beginning of
the amplification of the toroidal component of the magnetic field
to the moment of explosion is much shorter then in the results of
1D simulations. At the Figure \ref{texpl}, the dependence of the
time of the explosion (which is calculated as a time from the
beginning of the evolution of the magnetic field up to the moment
of the beginning of supernova explosion) from $\alpha$ value is
represented. For $\alpha>\sim 10^{-4}$ the $1/\sqrt{\alpha}$ law
holds approximately, but for smaller values of $\alpha$ the
explosion time $t_{explosion}$ is proportional
$t_{explosion}\sim\frac{1}{lg(\alpha)}$.

{\it Acknowledgements} G.S.B.-K. and S.G.M are grateful to the RFBR for the partial support by grant 02-02-16900,
NATO for the support in the frames of Collaborative Linkage Grant and The Royal Society for the support in the
frames of the international Joint Project. S.G.M. thanks the Organizers of the meeting for the support and hospitality.


\begin{thebibliography}{}

\bibitem[Akiyama et al.(2003)]{akiyama}
Akiyama S., Wheeler J.C., Meier D.L., Lichtenstadt I., 2003, ApJ, 584, 954

\bibitem[Ardeljan et al.(2004)]{abkkm}
Ardeljan N.V., Bisnovatyi-Kogan G.S., Kosmachevskii K.V., Moiseenko S.G.,
  2004, Astrophysics,  47, 37

\bibitem[Ardeljan et al.(2000)]{abkm}
Ardeljan N.V., Bisnovatyi-Kogan G.S., Moiseenko S.G.,
  2000, A\&A,  355, 1181

\bibitem[Ardeljan et al.(1979)]{abkp}
Ardeljan N.V., Bisnovatyi-Kogan G.S., Popov Yu.P.,  1979,
Astron. Zh. 56, 1244

\bibitem[Ardeljan \& Kosmachevskii (1995)]{arkos}
Ardeljan N.V., Kosmachevskii K.V.,
  1995, Comput. Math. Mod.,  6, 209

\bibitem[Bisnovatyi-Kogan(1970)]{bk1970}
Bisnovatyi-Kogan, G.S. 1970, \sovast, 47, 813

\bibitem[Bisnovatyi-Kogan et al.(1976)]{bk76}
    Bisnovatyi-Kogan, G.S., Popov, Iu.P., Samokhin, A.A.,
1976, Ap\&SS, 41, 287

\bibitem[Dungey(1958)]{dungey}
Dungey J.W. Cosmic electrodynamics. Cambridge Univ. Press, Cambridge,1958

\bibitem[Le Blanck \& Wilson(1970)]{lw1970}
Le Blanck, L.M. \& Wilson, J.R. 1970 \apj, 161, 541

\bibitem[Kotake et al.(2004)]{kotake}
Kotake K., Sawai H., Yamada S., Sato K., 2004, ApJ, 608, 391

\bibitem[Ohnishi(1983)]{ohnishi}
Ohnishu N., 1983, Tech. Rep. Inst. At. En. Kyoto Univ., No.198

\bibitem[Spruit(2002)]{spruit}
Spruit H.C., 2002, A\&A, 381, 923

\bibitem[Symbalisty(1984)]{symbalisty}
Symbalisty E.M.D., 1984, ApJ, 285, 729

\bibitem[Tayler(1973)]{tayler}
Tayler R.J., 1973, MNRAS, 161, 365

\bibitem[Yamada \& Sawai(2004)]{yamada}
Yamada S., Sawai H., 2004, ApJ, 608, 907

\end{thebibliography}
\end{document}